\documentclass[aps,prl,nofootinbib,reprint,longbibliography]{revtex4-2}

\usepackage{lipsum}
\usepackage{amsmath,amssymb,amsfonts,bbold}
\usepackage{braket}
\usepackage{natbib}
\usepackage{graphicx,subfigure}
\usepackage{hyperref}
\usepackage{mathtools}
\usepackage{comment}
\usepackage[normalem]{ulem}
\newcommand{\be}{\begin{equation}}
\newcommand{\ee}{\end{equation}}
\newcommand{\bse}{\begin{subequations}}
\newcommand{\ese}{\end{subequations}}
\newcommand{\ba}{\begin{eqnarray}}
\newcommand{\ea}{\end{eqnarray}}
\newcommand{\bea}{\begin{eqnarray}}
\newcommand{\eea}{\end{eqnarray}}

\newcommand{\lb}{\left (}
\newcommand{\rb}{\right )}
\newcommand{\mL}{\mathcal{L}}

\usepackage{color}
\usepackage[normalem]{ulem}  

\begin{document}


\title{Decoding the string in terms of holographic quantum maps}

\author{Avik Chakraborty}
\email{avik.phys88@gmail.com}
\affiliation{Departamento de Ciencias F\'isicas, Facultad de Ciencias Exactas, Universidad Andres Bello,
Sazi\'e 2212, Piso 7, Santiago, Chile}
\author{Tanay Kibe}
\email{tanay.kibe@wits.ac.za }
\affiliation{National Institute for Theoretical and Computational Sciences,
School of Physics and Mandelstam Institute for Theoretical Physics,
University of the Witwatersrand, Wits, 2050, South Africa}
\author{Mart\'in Molina}
\email{martinmolinaramos95@gmail.com}
\affiliation{Instituto de F\'{\i}sica, Pontificia Universidad Cat\'{o}lica de Valpara\'{\i}so,
Avenida Universidad 330, Valpara\'{\i}so, Chile}
\affiliation{Departamento de F\'{\i}sica, Universidad T\'{e}cnica Federico Santa Mar\'{\i}a,
Casilla 110-V, Valpara\'{\i}so, Chile,}
\author{Ayan Mukhopadhyay}
\email{ayan.mukhopadhyay@pucv.cl}
\affiliation{Instituto de F\'{\i}sica, Pontificia Universidad Cat\'{o}lica de Valpara\'{\i}so,
Avenida Universidad 330, Valpara\'{\i}so, Chile}
\author{Hardik Vamshi}
\email{ph22b009@smail.iitm.ac.in}
\affiliation{Department of Physics, Indian Institute of Technology Madras, Chennai 600036, India}

\date{\today}
\begin{abstract}
It has recently been shown that the Nambu-Goto equation for a string emerges from the junction conditions in three-dimensional gravity. Holographically, gravitational junctions are dual to interfaces in conformal field theory. We demonstrate at the level of linearized gravitational perturbations that each stringy mode of the junction corresponds to a $\mathcal{H}_{in}\rightarrow \mathcal{H}_{out}$ quantum map which can be factorized into a scattering matrix involving reflection/transmission and a relative automorphism of the Virasoro algebra, and also a $\mathcal{H}_{L}\rightarrow \mathcal{H}_{R}$ map of similar nature. These maps preserve the conformal boundary condition, are independent of the background conformal frame, as in the case of conformal interfaces studied in the literature, and realize a tunable energy transmitter.

\end{abstract}

\maketitle
\textit{Introduction:-} The holographic duality states that quantum gravity, particularly a superstring theory, can be reformulated in terms of a nongravitational quantum field theory living  at the boundary of spacetime \cite{Maldacena:1997re,Gubser:1998bc,Witten:1998qj}. There has been a lot of progress in our understanding of how the bulk spacetime and semi-classical bulk effective theory consisting of Einstein's gravity coupled to a few fields can be reconstructed from the boundary quantum field theory in the large N and strong coupling limit \cite{harlow2018tasi,Jahn:2021uqr,Chen:2021lnq,Kibe:2021gtw}. The reconstruction of \textit{dynamical} extended objects (branes) of the gravitational theory that are necessary for its non-perturbative completion, is therefore a problem of fundamental importance. 

Gravitational junctions between three-dimensional asymptotically anti-de Sitter (AdS$_3$) spacetimes model properties of conformal interfaces in dual two-dimensional conformal field theories (CFTs) \cite{Karch:2000ct,Karch:2001cw,DeWolfe:2001pq,Bachas:2001vj}, which appear also in many condensed matter systems, e.g. in quantum wire junctions \cite{Wong:1994np, Chamon:2003tz,Oshikawa:2005fh}, and defects in spin systems \cite{Oshikawa:1996dj} and quantum Hall systems \cite{fendley1995exact,fal1999topological,Gromov:2016umy}.  

The Nambu-Goto equation for the string has recently been shown to emerge from the gravitational junction conditions in three dimensions \cite{Banerjee:2024sqq}. In this letter, we demonstrate at the level of linearized gravitational perturbations how each stringy mode of a gravitational junction gluing two three-dimensional locally AdS$_3$ spacetimes can be decoded as a $\mathcal{H}_{in}\rightarrow \mathcal{H}_{out}$ quantum map from the in to out Hilbert spaces of excitations scattered at the dual interface. Equivalently, each stringy mode can be reformulated as a $\mathcal{H}_{L}\rightarrow \mathcal{H}_{R}$ quantum map relating the Hilbert spaces of the two dual CFTs that straddle the interface. These maps, which we explicitly construct in the universal sector in the large N limit, preserve the conformal boundary condition, and generalize those realized by conformal interfaces studied in the literature by involving relative automorphisms of the Virasoro algebra. We also prove that the quantum maps realized by the holographic defect operators corresponding to junctions with stringy excitations are independent of the conformal frame (the smoothened dual geometrical background spacetime) at the level of linearized departure from the vacuum (pure AdS$_3$). This reproduces and generalizes the universality of the reflection/transmission properties of conformal defect operators in CFT that are determined only by their boundary entropies \cite{Meineri:2019ycm}.

\textit{The conformal interface:-} The non-vanishing components of the energy-momentum tensor $T_{\mu\nu}$ in any state of a two-dimensional CFT are $T_{\pm\pm} = T_{\mu\nu}n_\pm^\mu n_\pm^\nu$, with $n_\pm^\mu$ being the two future directed null vectors. 
An interface is a gluing of two CFTs, namely CFT$_L$ and CFT$_R$ on the left and right, respectively. 
A conformal interface at $x=0$ is represented by an operator insertion $I_{L,R}$ which satisfies 
\begin{equation}\label{Eq:CBC1}
   (L_{n,+}^L -L_{-n,-}^L  )I_{L,R} = I_{L,R}(L_{n,+}^R -L_{-n,-}^R),
\end{equation}
with $L_{n,\pm}^{L}$ and $L_{n,\pm}^{R}$ being the Virasoro generators of CFT$_L$ and CFT$_R$, respectively, as $T_{++}-T_{--}$ generates conformal transformations that leave the line $x=0$ invariant. Equivalently, folding the right half of spacetime at $x=0$ and applying reflection on CFT$_R$, we can represent $I_{L,R}$ as a boundary state $\ket{B}$ in CFT$_L$ $\otimes$ $\overline{\rm CFT}_R$ satisfying \cite{Oshikawa:1996dj,Bachas:2001vj,Quella:2006de}
\begin{equation}\label{Eq:CBC2}
    (L_{n,+}^L  +L_{n,+}^R -L_{-n,-}^L-L_{-n,-}^R )\ket{B} =0.
\end{equation}
In either way, the interface represents a linear map $\mathcal{H}_{L}\rightarrow \mathcal{H}_{R}$ from states in CFT$_L$ to states in CFT$_R$.

\begin{figure}
    \centering
    \includegraphics[width=0.45\textwidth]{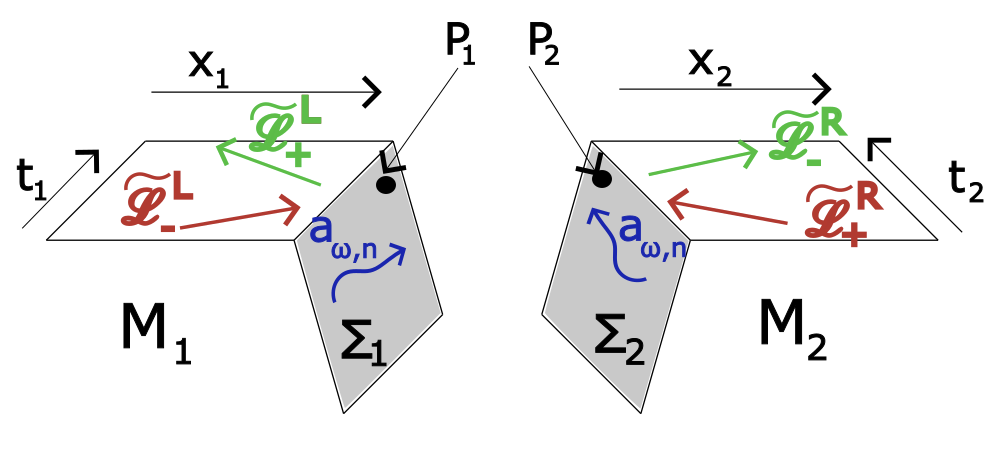}
    \caption{A two-way gravitational junction formed by gluing two AdS$_3$ manifolds by identifying points $P_{1,2}$ on the (gray) hypersurfaces. The incoming and outgoing plane wave amplitudes are indicated in the boundary CFT by arrows. The $a_{\omega,n}$ are classical string excitations.}
    \label{fig:setup}
\end{figure}

The interface can also be viewed as a $\mathcal{H}_{in}\rightarrow \mathcal{H}_{out}$ map, e.g. the $S$ matrix relating the incoming right-moving and left-moving energy excitations on the left and right sides of the interface, respectively, and the outgoing left-moving and right-moving energy excitations on the respective left and right sides (see Fig.~\ref{fig:setup}). When both CFTs have the same central charge $c$, this frequency independent $S$ matrix is
\begin{equation}\label{Eq:S}
   \begin{pmatrix} \langle T_{++}^L(\omega)\rangle\\ \langle T_{--}^R (\omega)\rangle\end{pmatrix} = S\begin{pmatrix}\langle T_{--}^L(\omega)\rangle\\\langle T_{++}^R(\omega)\rangle\end{pmatrix},\quad S =\begin{pmatrix}
   \frac{2}{2+\lambda}&\frac{\lambda}{2+\lambda}\\
    \frac{\lambda}{2+\lambda}&\frac{2}{2+\lambda}
\end{pmatrix},
\end{equation}
with $\lambda = \frac{2(c-c_{LR})}{c_{LR}}$, where $c_{LR}$ is the coefficient that appears in the two-point function $\langle T_{++}^L(x)T_{++}^R(y)\rangle_I$ in presence of the defect operator \cite{Meineri:2019ycm} and also determines the boundary entropy \cite{Azeyanagi:2007qj,Afxonidis:2024gne,Afxonidis:2025jph} ($T_{\pm\pm}^{L,R}(\omega)$ refers to the Fourier coefficients of $T_{\pm\pm}^{L,R}(x^\pm)$ with $x^\pm = t\pm x$). Note that $S$ is independent of the frequency in the absence of any dimensionful parameter in the CFTs glued at the interface.

Clearly, $S$ is simply the independent transmission/reflection of the incoming energies from the left and right with equal transmission coefficients
\begin{equation}
    \mathbb{T} = \frac{2}{2+\lambda},
\end{equation}
and equal reflection coefficients $\mathbb{R} = 1- \mathbb{T}$, implying energy conservation ($\langle T_{--}^L(\omega)\rangle+ \langle T_{++}^R (\omega)\rangle= \langle T_{--}^R (\omega)\rangle+ \langle T_{++}^L(\omega)\rangle$), which follows from the conformal boundary condition \eqref{Eq:CBC1}. Remarkably, $S$ is universal for the total energy fluxes, i.e. independent of how the energy fluxes are created \cite{Meineri:2019ycm}. Crucially, an overall (diagonal) conformal transformation on CFT$_L$ and CFT$_R$ does not affect $S$. The achronal ANEC (averaged null energy condition) implies that $c_{LR} \leq c$ \cite{Meineri:2019ycm}, i.e. $\lambda \geq 0$.

\textit{Gravitational two-way junction:-} The conformal interface is holographically dual to a gravitational two-way junction. Consider two three-dimensional locally AdS$_3$ manifolds $\mathcal{M}_{1,2}$. Each of these is split into two parts $\mathcal{M}_{i \alpha_i}$, $i=1,2$, $\alpha_i=L,R$, by co-dimension-1 hypersurfaces $\Sigma_{1,2}$. A gravitational junction $\Sigma$ involves the joining of one fragment each of $\mathcal{M}_1$ and $\mathcal{M}_2$. Here we will glue $\mathcal{M}_{1L}$ and $\mathcal{M}_{2R}$. The full spacetime $\widetilde{\mathcal{M}}$ is formed by the gluing of $\mathcal{M}_{1L}$ and $\mathcal{M}_{2R}$ at $\Sigma$ by identifying points on $\Sigma_{1,2}$, which are the images of $\Sigma$ in $\mathcal{M}_{1,2}$. This identification of points and the embeddings of $\Sigma_{1,2}$ in $\mathcal{M}_{1,2}$ should satisfy the gravitational junction conditions \cite{Israel:1966rt}.

Let $\mathcal{M}_{1,2}$ have the coordinates $t_{1,2},x_{1,2},z_{1,2}$, with  $x_{1,2}$ the coordinates transverse to $\Sigma_{1,2}$ and $z_{1,2}$ the radial coordinates. The embeddings of $\Sigma_{1,2}$ are specified by two functions $f_{1,2}$
\begin{equation}
    \Sigma_{1,2}: \quad x_{1,2}= f_{1,2}(t_{1,2},z_{1,2}).
\end{equation}
The freedom of choice of the coordinates of $\Sigma$ is fixed by defining the worldsheet coordinates $\tau$ and $\sigma$ at a point $P$ on $\Sigma$ as
\begin{equation}
    \tau(P)=\frac{t_1(P_1)+t_2(P_2)}{2},\quad  \sigma(P)=\frac{z_1(P_1)+z_2(P_2)}{2},
\end{equation}
where $P_{1,2}$ are the points on $\Sigma_{1,2}$ that are identified with $P$ (see Fig. \ref{fig:setup}). The following four variables, which are functions of $\tau$ and $\sigma$, completely specify the junction
\begin{equation}\label{Eq:vars}
    \tau_d = \frac{t_2-t_1}{2},\quad \sigma_d =\frac{z_2-z_1}{2},\quad x_{s,d}=\frac{f_2\pm f_1}{2}.
\end{equation}

The junction conditions can be obtained from the action
\begin{align}\label{Eq:bulk-action}
   & S_{grav}=\frac{1}{16\pi G_N}\int_{\mathcal{\widetilde{\mathcal{M}}}}d^3x \sqrt{-g}(R - 2\Lambda)+T_0\int_{\Sigma}{\rm d}\tau{\rm d}\sigma\,\sqrt{-\gamma}\nonumber\\&\quad+{\rm GHY \ terms},
\end{align}
where the metric is the \textit{only} degree of freedom. Above GHY are the Gibbons-Hawking-York (GHY) boundary terms and $T_0$ is the tension. This action is defined assuming that the induced metrics $\gamma_{1,2}$ on $\Sigma_{1,2}$ are identical, which defines the worldsheet metric as
\begin{equation}\label{Eq:metriccont}
    \gamma_{\mu\nu}(\tau,\sigma) =\gamma_{1,\mu\nu}(\tau,\sigma)= \gamma_{2,\mu\nu}(\tau,\sigma).
\end{equation}
Varying \eqref{Eq:bulk-action} away from the junction implies that the manifold is Einstein. At the junction, we have
\begin{equation}\label{Eq:Kdisc}
    \left(K_{i,\mu\nu} - K_i\,\gamma_{i,\mu\nu}\right) \vert_{\rm disc}= 8\pi G_N T_0 \gamma_{\mu\nu},
\end{equation}
where $K_{i,\mu\nu}$ is the extrinsic curvature of $\Sigma_i$ in $\mathcal{M}_{i\alpha_i}$, $K_i=K_{i,\mu\nu}\gamma^{\mu\nu}$ and $\vert_{\rm disc}$ denotes the discontinuity. The bulk diffeomorphism symmetry implies that the left hand side of \eqref{Eq:Kdisc} is conserved. We therefore obtain only one independent equation from \eqref{Eq:Kdisc}, which together with \eqref{Eq:metriccont} yields four equations for the four unknown functions \eqref{Eq:vars}.

It has been shown in \cite{Banerjee:2024sqq}, when $\mathcal{M}_1$ and $\mathcal{M}_2$ are copies of an Einstein manifold $\mathcal{M}$, that the general solutions of the junction conditions are in one-to-one correspondence with solutions of the non-linear Nambu-Goto equation for a worldsheet in $\mathcal{M}$. Specifically, the hypersurface $$\Sigma_{NG}: t=\tau, \quad z=\sigma, \quad x= x_s(\tau,\sigma),$$ corresponds to a solution of the non-linear Nambu-Goto equation in $\mathcal{M}$ when the tension $T_0$ vanishes, while $\tau_d$, $\sigma_d$ and $x_d$ are fixed completely by $x_s$ up to six rigid parameters related to worldsheet and spacetime isometries, which are irrelevant for the present paper. 

Here we will focus on locally AdS$_3$ Ba\~nados metrics \cite{Banados:1998gg}
\begin{multline}
    ds^2= \frac{dz^2}{z^2}+2dtdx\left(\mL_+(x^+)-\mL_-(x^-)\right)\\- \frac{dt^2}{z^2}\left(1-z^2\mL_+(x^+)\right)\left(1-z^2\mL_-(x^-)\right)\\+\frac{dx^2}{z^2}\left(1+z^2\mL_+(x^+)\right)\left(1+z^2\mL_-(x^-)\right),
\end{multline}
where $x^\pm=t\pm x$, and we have set $\Lambda = -1$. We assume that $\mathcal{M}_{1,2}$ have the above metrics with
\begin{equation}
    \mL_{\pm}^{(1)}(x_1^\pm)= \mL^{L}_{\omega,\pm} e^{i \omega x_{1}^\pm},\quad \mL^{(2)}_{\pm}(x_2^\pm)= \mL^{R}_{\omega,\pm} e^{i \omega x_{2}^\pm}.
\end{equation}
Furthermore, we assume that $\mL^{L,R}_{\omega,\pm}$ are small amplitudes of $\mathcal{O}(\epsilon)$. Our analysis will be linear and the above plane waves can be superposed to form wavepackets. At $\mathcal{O}(\epsilon^0)$ we have the exact solution
\begin{equation}\label{Eq:order0soln}
    \tau_d =0, \quad \sigma_d=0, \quad x_s=0, \quad x_d=-\frac{\lambda \sigma}{\sqrt{4-\lambda^2}},
\end{equation}
where $\lambda = 8\pi G_N T_0$ and $0\leq\lambda\leq 2$.
The equations \eqref{Eq:metriccont} and \eqref{Eq:Kdisc} can be solved perturbatively in $\epsilon$.
\begin{widetext}
 At $\mathcal{O}(\epsilon)$ we get the following equation
 \begin{equation}
     16 \sigma \partial^2_\tau x_s +4(4-\lambda^2)\lb2 \partial_\sigma x_s-\sigma\partial_\sigma^2x_s\rb =- i e^{i\omega \tau}(-4+\lambda^2)\sigma^3\omega\lb(\mL_{\omega,-}^{R}-\mL_{\omega,+}^{L})e^{\frac{i\lambda\omega\sigma}{\sqrt{4-\lambda^2}}}+(\mL_{\omega,-}^{L}-\mL_{\omega,+}^{R})e^{-\frac{i\lambda\omega \sigma}{\sqrt{4-\lambda^2}}}\rb,
     \label{Eq:xseq}
 \end{equation}
 which is the just the linearized Nambu-Goto (NG) equation in empty AdS with sources proportional to $(\mL_{\omega,-}^{L}-\mL_{\omega,+}^{R})$ and $(\mL_{\omega,-}^{R}-\mL_{\omega,+}^{L})$ when $\lambda \to 0$. This implies that the correspondence between the Nambu-Goto equation and junction conditions shown in \cite{Banerjee:2024sqq} generalizes (with sources) even when the backgrounds on both sides depart from each other. We obtain the following solution to \eqref{Eq:xseq}
\begin{align}
    x_s&=e^{i\omega \tau}\Bigg(\frac{-i(4-\lambda ^2)^{1/4} \left(\sin \left(\frac{2 \sigma \omega }{\sqrt{4-\lambda ^2}}\right) \left(A_{\omega,1} \sqrt{4-\lambda ^2}+2 A_{\omega,2} \sigma \omega \right)+\cos \left(\frac{2 \sigma \omega }{\sqrt{4-\lambda ^2}}\right) \left(A_{\omega,2} \sqrt{4-\lambda ^2}-2 A_{\omega,1} \sigma \omega \right)\right)}{2 \sqrt{\pi } (\omega )^{3/2}} \nonumber\\
    &\quad\quad\qquad+\frac{(\mL^{R}_{\omega,-}-\mL^{L}_{\omega,+}) e^{\frac{i \lambda  \sigma \omega }{\sqrt{4-\lambda ^2}}} \left(2 \sqrt{4-\lambda ^2} \lambda  \sigma \omega-i \left(\lambda ^2-4\right) \left(\sigma^2 \omega ^2+2\right) \right)}{4 \left(\lambda ^2-4\right) \omega ^3}\nonumber\\ &\quad\quad\qquad +\frac{(\mL^{L}_{\omega,-}-\mL^{R}_{\omega,+}) e^{-\frac{i \lambda  \sigma \omega }{\sqrt{4-\lambda ^2}}} \left(-2 \sqrt{4-\lambda ^2} \lambda  \sigma \omega-i \left(\lambda ^2-4\right) \left(\sigma^2 \omega ^2+2\right) \right)}{4 \left(\lambda ^2-4\right) \omega ^3}\Bigg)+\mathcal{O}(\epsilon^2),
\end{align}
where the first line is the solution of the source-free (homogeneous) equation \eqref{Eq:xseq}. Imposing ingoing boundary conditions at the Poincar\'{e} horizon \cite{Son:2002sd,Herzog:2002pc,Skenderis:2008dg} we obtain the coefficients $A_{\omega,1}=A_{\omega,nn}+A_{\omega,n}$ and $A_{\omega,2}=i A_{\omega,nn}$ which are of $\mathcal{O}(\epsilon)$. Here $A_{\omega,nn}$ corresponds to a non-normalizable mode of the homogeneous NG equation, which is the causal response to bulk perturbations that travels from the boundary towards the Poincar\'{e} horizon. $A_{\omega,n}$ is an intrinsic normalizable (stringy) mode of the homogeneous NG equation.  Both $A_{\omega,nn}$ and $A_{\omega,n}$ are determined by initial and boundary conditions as usual in Lorentzian holographic duality.  Explicit solutions for the other variables are in the End-matter.
\end{widetext}

The energy-momentum tensors on both sides of the dual CFT interface can be extracted using holographic renormalization \cite{Henningson:1998gx,Balasubramanian:1999re}. Explicitly, 
\begin{align}
    \braket{{T}^{L}_{\pm\pm}(\omega)} = \frac{c}{12\pi} \mL_{\omega,\pm}^{L},\,\,
    \braket{{T}^{R}_{\pm\pm}(\omega)} = \frac{c}{12\pi} \mL_{\omega,\pm}^{R} \label{Eq:energy}.
\end{align}

The Dirichlet boundary conditions corresponding to the interface at $x_1 =x_2 =0$ impose $\lim_{\sigma\to0}x_d= \lim_{\sigma\to0}x_s =0$. From $\lim_{\sigma\to0}x_d =0$ we obtain
\begin{equation}\label{Eq:dir4}
    \mL_{\omega,-}^{L}+\mL_{\omega,+}^{R}=\mL_{\omega,-}^{R}+\mL_{\omega,+}^{L}.
\end{equation}
This is the conservation of energy at the interface, or equivalently the conformal boundary condition. Eq. \eqref{Eq:dir4} can be solved by
\begin{align}\label{Eq:dir4sol}
    &\mL_{\omega,+}^{L}=\mathcal{T}_{\omega,L} \mL_{\omega,+}^{R}+(1-\mathcal{T}_{\omega,R}) \mL_{\omega,-}^{L},\nonumber\\
    &\mL_{\omega,-}^{R}=\mathcal{T}_{\omega,R} \mL_{\omega,-}^{L}+(1-\mathcal{T}_{\omega,L})\mL_{\omega,+}^{R},
\end{align}
where $\mathcal{T}_{\omega,L(R)}$ and $1-\mathcal{T}_{\omega,L(R)}$ are arbitrary (frequency dependent) transmission and reflection coefficients for the left and right-movers, respectively (note these are $\mathcal{O}(\epsilon^0)$). From $\lim_{\sigma\to0}x_s =0$, we obtain that
\begin{equation}
    A_{\omega,nn} = i\frac{2 \sqrt{\pi } ( \mL_{\omega,-}^{L} \mathcal{T}_{\omega,R} -\mL_{\omega,+}^{R} \mathcal{T}_{\omega,L})}{\left(4-\lambda ^2\right)^{3/4} \omega ^{3/2}}.
\end{equation}

\textit{The string as a holographic quantum map:-} In \cite{Bachas:2020yxv}, it was shown that the transmission/reflection coefficients of the CFT interface can be derived by keeping the incoming energy flux only on one side, considering the causal response on the worldsheet with $A_{\omega,n} = 0$ and using the Dirichlet boundary conditions on the other variables $\tau_d$ and $\sigma_d$. However, the gravitational problem is fundamentally non-linear, and therefore assuming independent and equal transmission from the left and right sides affect the higher-order perturbative expansions. Furthermore, as discussed below, setting Dirichlet boundary conditions for $\tau_d$ and $\sigma_d$ is inadequate for proving that the scattering process is independent of the conformal frame, i.e. the smoothened dual background geometry in the absence of the interface (which is assumed to be the empty AdS$_3$ space dual to the CFT vacuum on both sides).

The key point is that, instead of setting Dirichlet boundary conditions for $\tau_d$ and $\sigma_d$, we should use relative conformal transformations to obtain continuous coordinates and metric across the dual interface at the boundary in which physical experiments are set up. Let us define
\begin{equation}
    \lim_{\sigma\rightarrow0}\tau_d(\tau,\sigma)=\mathbb{t}_{\omega,d}e^{i\omega\tau}+ \mathcal{O}(\epsilon^2)
\end{equation}
so that at the boundary $t_{1,2}(\tau) =\tau\mp\mathbb{t}_{\omega,d}e^{i\omega\tau}+ \mathcal{O}(\epsilon^2)$. In agreement with \cite{Banerjee:2024sqq}, we find that the relative time reparametrization $\mathbb{t}_d(\tau)$ encodes the normalizable stringy mode as both $\mathbb{t}_{\omega,d}$ and $\lim_{\sigma\to0} \frac{\sigma_d(\tau,\sigma)}{\sigma}$ are proportional to
\begin{multline*}
    -\mL_{\omega,-}^{L}(-2+\mathcal{T}_{\omega,R}(2+\lambda))+ \mL_{\omega,+}^{R}(-2+\mathcal{T}_{\omega,L}(2+\lambda))+ 2a_{\omega,n},
\end{multline*}
where $a_{\omega,n}=\frac{iA_{\omega,n}\lambda(4-\lambda^2)^{3/4}\omega^{3/2}}{4\sqrt{\pi}}$.

The discontinuity in the time coordinates ($t_1$ and $t_2$) at the interface located at $x_1 = x_2=0$  can be undone using separate conformal transformations on the two sides which involve coordinate transformations
\begin{align}\label{Eq:CT}
    &\widetilde{t}_{1,2} =\frac{1}{2}(\mathbb{h}_{1,2}^{-1}(t_{1,2}+x_{1,2})+\mathbb{h}^{-1}_{1,2}(t_{1,2}-x_{1,2})),\nonumber\\
    &\widetilde{x}_{1,2} =\frac{1}{2}(\mathbb{h}_{1,2}^{-1}(t_{1,2}+x_{1,2})-\mathbb{h}_{1,2}^{-1}(t_{1,2}-x_{1,2})),
\end{align}
and the associated Weyl transformations that brings the metric on both sides back to the Minkowski form. The most general choices of $\mathbb{h}_{1,2}$ at the linear order are
\begin{align}\label{Eq:h12}
    \mathbb{h}_2(\tau) = \tau+  \kappa_\omega \mathbb{t}_{\omega,d}e^{i\omega\tau},\quad  \mathbb{h}_1(\tau) = \tau+ (\kappa_\omega-2) \mathbb{t}_{\omega,d}e^{i\omega\tau},
\end{align}
where $\kappa_\omega$ is an arbitrary $\mathcal{O}(\epsilon^0)$ (frequency dependent) parameter which acts as an overall (diagonal) conformal transformation that changes the conformal frame (see below). It is easy to see using \eqref{Eq:h12} that \eqref{Eq:CT} preserves the interface at $\tilde{x}_{1,2}=0$ where $\tilde{t}_2=\tilde{t}_1$. We therefore obtain continuous coordinates and metric across the interface. These conformal transformations can be uplifted to bulk diffeomorphisms ($X^\mu(\tau,\sigma) \rightarrow \widetilde{X}^\mu(\tau,\sigma))$ with $X^\mu$ denoting bulk coordinates on both sides \cite{Balasubramanian:1999re,Skenderis:2008dg,deHaro:2000vlm}, and both the induced metric and the extrinsic curvatures of $\Sigma_{1,2}$ remain invariant under these bulk diffeomorphisms, producing an equivalent solution of the junction conditions.

Under the conformal transformations, the energy-momentum tensors on both sides of the interface transform as
\begin{align}
    \widetilde{T}^{L,R}_{\pm\pm}(\tilde{x}^\pm)&= \mathbb{h}'_{1,2}(\widetilde{x}^\pm)^2 T_{\pm\pm}^{L,R}(\mathbb{h}_{1,2}(\widetilde{x}_2^\pm))\nonumber\\&\qquad-\frac{c}{24\pi}{\rm Sch}(\mathbb{h}_{1,2}(\tilde{x}^\pm),\tilde{x}^\pm)\nonumber\\
    &=\frac{c}{12\pi}e^{i \omega \tilde{x}_2^\pm}\tilde{\mL}_{\omega,\pm}^{L,R}+ \mathcal{O}(\epsilon^2), \label{Eq:newenergy}
\end{align}
where
\begin{align}
        &\tilde{\mL}_{\omega,+}^{L}=\frac{\mL_{\omega,+}^{R}}{4}\lb4+2(-1+\mathcal{T}_{\omega,L})\kappa_\omega +\mathcal{T}_{\omega,L}(-2+\kappa_\omega)\lambda\rb \nonumber\\
    & \quad +\frac{\mL_{\omega,-}^L}{4}\lb2\kappa_\omega+\mathcal{T}_{\omega,R}(2\lambda-\kappa_\omega(2+\lambda))\rb+\frac{a_{\omega,n}(\kappa_\omega-2)}{2},\label{Eq:newenergyLp}\\
    &\tilde{\mL}_{\omega,-}^{L}=\frac{\mL_{\omega,+}^{R}}{4}(\kappa_\omega-2)(-2+\mathcal{T}_{\omega,L}(2+\lambda)) \nonumber\\
    &  +\frac{\mL_{\omega,-}^L}{4}\lb2\kappa_\omega-\mathcal{T}_{\omega,R}(-2+\kappa_\omega)(2+\lambda)\rb+\frac{a_{\omega,n}(\kappa_\omega-2)}{2}, \label{Eq:newenergyLm}\\
    &\tilde{\mL}_{\omega,+}^{R}=\frac{\mL_{\omega,+}^{R}}{4}\lb4+\kappa_\omega(-2+\mathcal{T}_{\omega,L}(2+\lambda))\rb \nonumber\\
    & \quad -\frac{\mL_{\omega,-}^L}{4}\kappa_\omega\lb-2+\mathcal{T}_{\omega,R}(2+\lambda)\rb+\frac{\kappa_\omega a_{\omega,n}}{2}, \label{Eq:newenergyRp} \\
    &\tilde{\mL}_{\omega,-}^{R}=\frac{\mL_{\omega,+}^{R}}{4}\lb2(-1+\mathcal{T}_{\omega,L})(-2+\kappa_\omega)+\mathcal{T}_{\omega,L}\kappa_\omega\lambda\rb \nonumber\\
    &  +\frac{\mL_{\omega,-}^L}{4}\lb2\kappa_\omega +\mathcal{T}_{\omega,R}(4-\kappa_\omega(2+\lambda))\rb+\frac{\kappa_\omega a_{\omega,n}}{2}.\label{Eq:newenergyRm}
\end{align}
These transformations are reproduced in the bulk via the holographic renormalization procedure after implementing the bulk diffeomorphisms which uplift the corresponding conformal transformations \cite{deHaro:2000vlm}. The transformed amplitudes also satisfy energy conservation (and thus also conformal boundary condition) as
\begin{equation}\label{Eq:energycon}
    \tilde{\mL}_{\omega,-}^{L}+\tilde{\mL}_{\omega,+}^{R}=\tilde{\mL}_{\omega,-}^{R} +\tilde{\mL}_{\omega,+}^{L}.
\end{equation}

As evident from \eqref{Eq:newenergyLp}, \eqref{Eq:newenergyLm}, \eqref{Eq:newenergyRp} and \eqref{Eq:newenergyRm}, the four physical energy fluxes $\tilde{\mL}_{\omega,\pm}^{L,R}$ depend on the six parameters $\kappa_\omega$, $a_{\omega,n}$, $\mL_{\omega,+}^{R}$, $\mL_{\omega,-}^L$, $\mathcal{T}_{\omega,L}$ and $\mathcal{T}_{\omega,R}$. Remarkably, for arbitrary values of the six parameters, we obtain that the physical excitations satisfy
\begin{align}\label{Eq:HLHR1}
   \tilde{\mL}_{\omega,+}^{L}&=\frac{2}{2+\lambda}\lb \tilde{\mL}_{\omega,+}^R-\frac{2a_{\omega,n}}{2-\lambda}\rb + \frac{\lambda}{2+\lambda}\lb\tilde{\mL}_{\omega,-}^L+\frac{2a_{\omega,n}}{2-\lambda}\rb, \nonumber\\
    \tilde{\mL}_{\omega,-}^{R}&=\frac{\lambda}{2+\lambda}\lb\tilde{\mL}_{\omega,+}^R-\frac{2a_{\omega,n}}{2-\lambda}\rb+\frac{2}{2+\lambda}\lb \tilde{\mL}_{\omega,-}^L+\frac{2a_{\omega,n}}{2-\lambda}\rb,
\end{align}
This implies that the dual interface acts as a $\mathcal{H}_{in}\rightarrow\mathcal{H}_{out}$ quantum map in the universal sector given by $\mathcal{S}\circ\mathcal{D}$ where $S$ is the matrix \eqref{Eq:S} and $\mathcal{D}$ is a redistribution of energy among the two incoming energy excitations, which can be realized by the linearized conformal transformation given by
\begin{equation}\label{Eq:fexp}
    g_\pm(x^{\pm}) = x^\pm \mp  \frac{e^{i \omega x^\pm}}{\omega^3}\frac{4 i a_{\omega,n}}{2-\lambda} +\mathcal{O}(\epsilon^2).
\end{equation}
Crucially, the quantum map $\mathcal{S}\circ\mathcal{D}$ is independent of $\kappa_\omega$ which changes all $\tilde{\mL}_{\omega,\pm}^{L,R}$ by an additive factor of $\kappa_\omega a_{\omega,n}/2$ and thus the overall conformal frame (i.e. the smoothened background geometry) reproducing and generalizing this feature of static conformal interfaces \cite{Meineri:2019ycm}. The parameter $a_{\omega,n}$ determining the energy redistribution $\mathcal{D}$ can be extracted from $\langle D(\omega) \rangle$, the expectation value of the displacement operator (which is also independent of $\kappa_\omega$) as shown using the Ward identities (see Supplemental material \cite{supp}, which includes Ref.\cite{Bianchi:2015liz}, for a derivation) . 

The physical excitations \eqref{Eq:newenergyLp}, \eqref{Eq:newenergyLm}, \eqref{Eq:newenergyRp} and \eqref{Eq:newenergyRm} also satisfy
\begin{align}\label{Eq:HLHR2}
   \tilde{\mL}_{\omega,+}^{L}&=\frac{2}{2+\lambda} \tilde{\mL}_{\omega,+}^R + \frac{\lambda}{2+\lambda}\tilde{\mL}_{\omega,-}^L - \frac{2 a_{\omega,n}}{2+\lambda}, \nonumber\\
    \tilde{\mL}_{\omega,-}^{R}&=\frac{\lambda}{2+\lambda}\tilde{\mL}_{\omega,+}^R+\frac{2}{2+\lambda}\tilde{\mL}_{\omega,-}^L + \frac{2 a_{\omega,n}}{2+\lambda},
\end{align}
so that the $\mathcal{H}_{in}\rightarrow\mathcal{H}_{out}$ map can be rewritten in the form ${\mathcal{D}}\circ\mathcal{S}$ where ${\mathcal{D}}$ is the same redistribution of 
energy realized via a linearized conformal transformation as discussed above acting on the \textit{outgoing} energy modes.

We also note from \eqref{Eq:newenergyLp}, \eqref{Eq:newenergyLm}, \eqref{Eq:newenergyRp} and \eqref{Eq:newenergyRm} that 
\begin{align}\label{Eq:HLHR}
   \tilde{\mL}_{\omega,+}^{R}&=\lb 1+\frac{\lambda}{2}\rb \tilde{\mL}_{\omega,+}^L -\frac{\lambda}{2}\tilde{\mL}_{\omega,-}^L + a_{\omega,n}, \nonumber\\
    \tilde{\mL}_{\omega,-}^{R}&= \frac{\lambda}{2} \tilde{\mL}_{\omega,+}^L +\lb 1-\frac{\lambda}{2}\rb\tilde{\mL}_{\omega,-}^L +a_{\omega,n}.
\end{align}
Thus the interface acts as a universal $\mathcal{H}_L \rightarrow\mathcal{H}_R$ map of the form $\mathcal{C}\circ\widetilde{\mathcal{S}}$  mapping the energy fluxes on left side to those on the right side with 
\begin{equation}
\widetilde{\mathcal{S}} =\begin{pmatrix}
    1+\frac{\lambda}{2}&-\frac{\lambda}{2}\\
    \frac{\lambda}{2}&1-\frac{\lambda}{2}
\end{pmatrix}
\end{equation}
being an invertible matrix and $\mathcal{C}$ is a conformal transformation, determined by $a_{\omega,n}$,  on the right side that acts in the same way on the left and right movers. 

For any value of $\lambda$ between $0$ and $2$, the interface is topological when 
\begin{equation}
    a_{\omega,n} =\frac{\lambda}{2}\lb\tilde{\mL}_{\omega,-}^L - \tilde{\mL}_{\omega,+}^R \rb.
\end{equation}
In this case, as evident from \eqref{Eq:HLHR1}, $\tilde{\mL}_{\omega,\pm}^R= \tilde{\mL}_{\omega,\pm}^L$ so that we obtain perfect energy transmission.  Similarly, the interface is always factorizing when
\begin{equation}
    a_{\omega,n} =\tilde{\mL}_{\omega,+}^R - \tilde{\mL}_{\omega,-}^L 
\end{equation}
as in this case $\tilde{\mL}_{\omega,+}^{R,L} = \tilde{\mL}_{\omega,-}^{R,L}$ and we obtain perfect energy reflection. Thus remarkably, the stringy interface is completely tunable. As discussed in the supplemental material \cite{supp}, the averaged null energy condition (ANEC) is not violated as the additional conformal transformation factors do not affect the total energy fluxes.

\paragraph{Generalized conformal interfaces:-} Our calculations up to the linear order in the energy excitations reveal that the holographic interface operator corresponding to a stringy junction mode is a modification of \eqref{Eq:CBC1} taking the form
\begin{equation}\label{Eq:CBC3}
   (L_{n,+}^L -L_{-n,-}^L  )I_{L,R} = I_{L,R}(\tilde{L}_{n,+}^R -\tilde{L}_{-n,-}^R),
\end{equation}
implying \eqref{Eq:energycon}, where $\tilde{L}_{m,\pm}$ are automorphisms of $L_{\pm,m}$ which preserve the Virasoro algebra. Although we have not computed the gravitational perturbations beyond the linearized order, our result addresses a key conceptual issue relevant even for the static holographic junction studied in \cite{Bachas:2020yxv} at the linearized level. Non-linearity of gravity, particularly the junction conditions, implies mixing of $\mL_{2\omega,\pm}^{L,R}$ with ${\mL_{\omega,\pm}^{L,R}}^2$ for instance, while the dual interface should realize a linear quantum map. Consider an automorphism of the form $L_n \rightarrow \tilde{L}_n= VL_n V^{-1}$ with
\[
V= \exp\left(\sum_m \zeta_m L_m +\sum_{k=2}^\infty\sum_{p_1}\cdots\sum_{p_k} \tilde{\zeta}_{p_1,\cdots p_k}L_{p_1}\cdots L_{p_k} \right).
\]
When $\tilde{\zeta}_{p_1,\cdots p_k}$ vanish, the above is a (non-)unitary conformal transformation which acts linearly on $L_n$. However, generally the automorphism act non-linearly on $L_n$ inspite of being a linear quantum map. So \eqref{Eq:CBC3} is consistent with the non-linearity of the gravitational problem.

At the level of linearized conformal transformations, the factors $\mathcal{D}$ and $\mathcal{C}$ in the quantum maps discussed above are conformal transformations determined only by the homogeneous Nambu-Goto modes irrespective of the background conformal frame. Given that the energy reflection/transmission properties of the dual interface are tunable, it would be of great importance to study how \eqref{Eq:CBC3} is realized for the dual interface both in case of the static holographic junction and the holographic junction corresponding to a non-trivial solution of the non-linear Nambu-Goto equation when the full non-linear gravitational perturbations are included.

\textit{Conclusions:-} In this letter, we have demonstrated that the gravitational two-way junction including its stringy degrees of freedom can be decoded in terms of quantum maps at the level of linearized gravitational perturbations.
In the absence of stringy vibrations, these maps reduce to frequency independent universal maps \cite{Quella:2006de,Kimura:2015nka,Billo:2016cpy,Meineri:2019ycm} that are realized by a defect operator, while the stringy modes correspond to generalized maps involving compositions with relative conformal transformations. We have particularly improved the methodology of \cite{Bachas:2020yxv} to show that the transmission and reflection on both sides of the interface occur independently, and the full quantum maps are independent of the conformal frame as can be anticipated from the literature \cite{Meineri:2019ycm}. Our results imply generalization of the conformal interfaces studied in \cite{Quella:2006de,Kimura:2015nka,Billo:2016cpy,Meineri:2019ycm} realizing tunable energy transmitters. We have argued that the generalized quantum maps are relevant even for the static holographic interface when the full non-linear gravitational perturbations are included.

It would be also of profound interest to understand how explicit reconstruction of sub-regions of the gravitational junction from interface CFT sub-regions can work by generalizing the reformulation of bulk reconstruction in holography in terms of recovery maps of quantum error correcting codes \cite{harlow2018tasi,Kibe:2021gtw}.

It would also be interesting to investigate the quantum null energy condition (QNEC) \cite{Bousso:2015mna,Wall:2017blw,Iizuka:2025xnd} in holographic interfaces dual to gravitational junctions. Understanding how the link between quantum energy conditions and quantum thermodynamic bounds demonstrated in \cite{Kibe:2021qjy,Banerjee:2022dgv,Kibe:2024icu,Kibe:2025cqc} (implying generalized Clausius inequalities for instance) are realized in interfaces is fundamental for actualization of applicable quantum devices, tunable transmitters for instance, via interfaces in many-body systems.

\begin{acknowledgments}
\textit{Acknowledgments:-}  We thank Costas Bachas, Avik Banerjee and Giuseppe Policastro for valuable discussions. We especially thank Giuseppe Policastro for discussions related to the ANEC bounds and quantum maps. AC, AM and MM acknowledge support from FONDECYT postdoctoral grant no. 3230222, FONDECYT regular grant no. 1240955 and ``Doctorado Nacional'' grant no. 21250596 of La Agencia Nacional de Investigaci\'{o}n y Desarrollo (ANID), Chile, respectively. AM gratefully acknowledges the hospitality of LPENS, where a substantial part of this work was carried out during his tenure as a CNRS invited professor.
\end{acknowledgments}

\newpage
\section{Appendix A: Solutions of the junction condition:} The solutions of the junction conditions for $\sigma_d$, $\tau_d$ and $x_d$ at the linear order are given below. Note that in the following solutions we have turned off additional terms that solve the homogeneous equations, since they are not of the plane-wave form and do not affect the scattering.
\begin{widetext}
\begin{align}
    &\sigma_d= \sigma e^{i\omega\tau}\left(\frac{(\mL^{L}_{\omega,-}-\mL^{R}_{\omega,+}) e^{-i\omega \sigma p_\lambda }+(\mL^{L}_{\omega,+}-\mL^{R}_{\omega,-}) e^{i\omega\sigma p_\lambda}}{2\omega^2}+\frac{i \lambda  \left(4-\lambda ^2\right)^{3/4} \left(A_{\omega,1} \cos \left(\frac{2 \sigma \omega }{\sqrt{4-\lambda ^2}}\right)-A_{\omega,2} \sin \left(\frac{2 \sigma \omega }{\sqrt{4-\lambda ^2}}\right)\right)}{4 \sqrt{\pi } \sqrt{\omega }}\right),\\
    &\tau_d=e^{i \omega \tau}\lb i\frac{(\mL^{L}_{\omega,-}-\mL^{R}_{\omega,+}) e^{-i\omega \sigma p_\lambda }+(\mL^{L}_{\omega,+}-\mL^{R}_{\omega,-}) e^{i\omega \sigma p_\lambda }}{4\omega^3(-2+\sigma^2\omega^2)^{-1}}+\frac{\lambda  \left(4-\lambda ^2\right)^{3/4} \left(A_{\omega,1} \cos \left(\frac{2 \sigma \omega }{\sqrt{4-\lambda ^2}}\right)-A_{\omega,2} \sin \left(\frac{2 \sigma \omega }{\sqrt{4-\lambda ^2}}\right)\right)}{4 \sqrt{\pi } \omega^{3/2}}\rb,
\end{align}
up to $\mathcal{O}(\epsilon^2)$ and $x_d=-\frac{\lambda \sigma}{\sqrt{4-\lambda^2}} + x_d^{(1)} + \mathcal{O}(\epsilon^2)$, with
\begin{equation}
    x_d^{(1)}= e^{i\omega\tau}\left(\frac{-2\lambda\sigma\left((\mL^{L}_{\omega,-}+\mL^{R}_{\omega,+}) e^{-i\omega\sigma p_\lambda}+(\mL^{L}_{\omega,+}+\mL^{R}_{\omega,-}) e^{i\omega\sigma p_\lambda}\right)}{\sqrt{4-\lambda^2}\omega^2}+\frac{i\left((\mL^{L}_{\omega,-}+\mL^{R}_{\omega,+}) e^{-i\omega\sigma p_\lambda}-(\mL^{L}_{\omega,+}+\mL^{R}_{\omega,-}) e^{i\omega\sigma p_\lambda}\right)}{\omega^3(2+\sigma^2 \omega^2)^{-1}}\right),
\end{equation}
where $p_\lambda=\frac{\lambda}{\sqrt{4-\lambda^2}}$.
\end{widetext}

\section{Appendix B: Ward identities at the interface and the displacement operator} 
To see the Ward identities we adopt continuous coordinates $\tilde{t},\tilde{x}$ where the energy-momentum tensors are
\begin{equation}
    \braket{\widetilde{T}^{L,R}_{\pm\pm}(\omega)} = \frac{c\epsilon}{12\pi}\tilde{\mL}_{\omega,\pm}^{L,R}+ \mathcal{O}(\epsilon^2)
\end{equation}
The full energy-momentum tensor can be written as
\begin{equation}\label{Eq:Tfull}
    \braket{\widetilde{T}_{\pm\pm}(\tilde{t},\tilde{x})} = \theta(-\tilde{x}) \braket{\widetilde{T}^L_{\pm\pm}(\tilde{x}^\pm)}+\theta(\tilde{x})\braket{\widetilde{T}^R_{\pm\pm}(\tilde{x}^\pm)},
\end{equation}
where $\theta(x)$ is the Heaviside theta function. We then have the following Ward identities
\begin{align}
    &\partial_{\tilde{t}}\braket{\widetilde{T}^{\tilde{t}\tilde{t}}}+\partial_{\tilde{x}}\braket{\widetilde{T}^{\tilde{x}\tilde{t}}}=0,\label{Eq:WI1}\\
    &\partial_{\tilde{t}}\braket{\widetilde{T}^{\tilde{t}\tilde{x}}}+\partial_{\tilde{x}}\braket{\widetilde{T}^{\tilde{x}\tilde{x}}}=\delta(\tilde{x})q(\tilde{t}),\label{Eq:WI2}
\end{align}
with the source 
\begin{align}
    q(\tilde{t})&={\braket{\widetilde{T}^R_{++}(\tilde{t})+\widetilde{T}^L_{++}(\tilde{t})-\widetilde{T}^R_{--}(\tilde{t})-\widetilde{T}^L_{--}(\tilde{t})}} +\mathcal{O}(\epsilon^2),\nonumber\\
                &= \frac{c e^{i\omega \tilde{t}}}{12\pi}\lb\tilde{\mL}_{\omega,+}^R+\tilde{\mL}_{\omega,+}^L-\tilde{\mL}_{\omega,-}^R-\tilde{\mL}_{\omega,-}^L\rb+\mathcal{O}(\epsilon^2), \nonumber\\
                &= \frac{c e^{i\omega \tilde{t}}}{12\pi}\frac{4}{2+\lambda}\lb\tilde{\mL}_{\omega,+}^R - \tilde{\mL}_{\omega,-}^L -a_{\omega,n}\rb+\mathcal{O}(\epsilon^2). \label{Eq:qt}
\end{align}
The Nambu-Goto mode $a_{\omega,n}$ appears in the source for the second Ward identity \eqref{Eq:WI2}. The source for the first identity \eqref{Eq:WI1} vanishes because of energy conservation at the interface, while the source $q(t)$ for the second Ward identity is the expectation value of the displacement operator $D$ (i.e. $q(t) =\langle D(t)\rangle)$, where
\begin{equation}
    D(t) =2 \lb \widetilde{T}_{++}^R - \widetilde{T}_{++}^L \rb = \lb \widetilde{T}_{++}^R - \widetilde{T}_{++}^L+ \widetilde{T}_{--}^R - \widetilde{T}_{--}^L\rb\,,
\end{equation}
with $\widetilde{T}_{++}^{L,R}$ evaluated at $x=0$. The second equality above follows from energy conservation. The displacement operator quantifies the energy cost of a small displacement of the interface \cite{Bianchi:2015liz}. 

We readily note that
\begin{equation}
    a_{\omega,n} =\frac{12\pi}{c}\left(\langle \widetilde{T}_{++}^R(\omega) - \widetilde{T}_{--}^L(\omega)  - \frac{2+\lambda}{4}D(\omega)\rangle\right)
\end{equation}
implying that the Nambu-Goto mode $a_{\omega,n}$ can be obtained from the expectation value of the displacement operator.

\section{Appendix C: Checking achronal ANEC}
Finally, the achronal ANEC is always satisfied for the generalized quantum maps at the linear order as the linearized conformal transformations 
\begin{equation}
    g_\pm(x^{\pm}) = x^\pm \mp  \frac{e^{i \omega x^\pm}}{\omega^3}\frac{4 i a_{\omega,n}}{2-\lambda} +\mathcal{O}(\epsilon^2)
\end{equation}
do not change the energy fluxes if $a_{\omega,n}(\omega)$ vanishes as $\omega \rightarrow 0$. Under the following transformation
\begin{equation}
    g(x^\pm) = x^\pm + \epsilon \mathbb{g}(x^\pm),
\end{equation}
the averaged null energy
 \begin{equation}
     \int{\rm d}x^\pm T_{\pm\pm}
 \end{equation} 
 is invariant at $\mathcal{O}(\epsilon)$ if 
 \begin{equation}
     \int{\rm d}x^\pm \mathbb{g}'''(x^+) =0,
 \end{equation} 
 when $T_{\pm\pm}$ are also $\mathcal{O}(\epsilon)$. For $\mathbb{t}_{\omega,d}$ and other variables to be well defined, we actually require that $a_{\omega,n}(\omega)$ vanishes at least as $\omega^3$ as $\omega \rightarrow 0$.

\bibliographystyle{apsrev4-1}
\bibliography{references}

@ARTICLE{Banerjee:2024sqq,
    author = "Banerjee, Avik and Mukhopadhyay, Ayan and Policastro, Giuseppe",
    title = "{Nambu-Goto equation from three-dimensional gravity}",
    eprint = "2404.02149",
    archivePrefix = "arXiv",
    primaryClass = "hep-th",
    doi = "10.1007/JHEP09(2024)013",
    journal = "JHEP",
    volume = "09",
    pages = "013",
    year = "2024"
}

@article{Kibe:2021qjy,
    author = "Kibe, Tanay and Mukhopadhyay, Ayan and Roy, Pratik",
    title = "{Quantum Thermodynamics of Holographic Quenches and Bounds on the Growth of Entanglement from the Quantum Null Energy Condition}",
    eprint = "2109.09914",
    archivePrefix = "arXiv",
    primaryClass = "hep-th",
    doi = "10.1103/PhysRevLett.128.191602",
    journal = "Phys. Rev. Lett.",
    volume = "128",
    number = "19",
    pages = "191602",
    year = "2022"
}

@article{Banerjee:2022dgv,
    author = "Banerjee, Avik and Kibe, Tanay and Mittal, Nehal and Mukhopadhyay, Ayan and Roy, Pratik",
    title = "{Erasure Tolerant Quantum Memory and the Quantum Null Energy Condition in Holographic Systems}",
    eprint = "2202.00022",
    archivePrefix = "arXiv",
    primaryClass = "hep-th",
    doi = "10.1103/PhysRevLett.129.191601",
    journal = "Phys. Rev. Lett.",
    volume = "129",
    number = "19",
    pages = "191601",
    year = "2022"
}

@article{Kibe:2024icu,
    author = "Kibe, Tanay and Mukhopadhyay, Ayan and Roy, Pratik",
    title = "{Generalized Clausius inequalities and entanglement production in holographic two-dimensional CFTs}",
    eprint = "2412.13256",
    archivePrefix = "arXiv",
    primaryClass = "hep-th",
    doi = "10.1007/JHEP04(2025)096",
    journal = "JHEP",
    volume = "04",
    pages = "096",
    year = "2025"
}

@article{Maldacena:1997re,
	archiveprefix = {arXiv},
	author = {Maldacena, Juan Martin},
	doi = {10.1023/A:1026654312961},
	eprint = {hep-th/9711200},
	journal = {Adv. Theor. Math. Phys.},
	pages = {231--252},
	reportnumber = {HUTP-97-A097, HUTP-98-A097},
	title = {{The Large N limit of superconformal field theories and supergravity}},
	volume = {2},
	year = {1998}}

@article{Gubser:1998bc,
	archiveprefix = {arXiv},
	author = {Gubser, S. S. and Klebanov, Igor R. and Polyakov, Alexander M.},
	doi = {10.1016/S0370-2693(98)00377-3},
	eprint = {hep-th/9802109},
	journal = {Phys. Lett. B},
	pages = {105--114},
	reportnumber = {PUPT-1767},
	title = {{Gauge theory correlators from noncritical string theory}},
	volume = {428},
	year = {1998}}

@article{Witten:1998qj,
	archiveprefix = {arXiv},
	author = {Witten, Edward},
	doi = {10.4310/ATMP.1998.v2.n2.a2},
	eprint = {hep-th/9802150},
	journal = {Adv. Theor. Math. Phys.},
	pages = {253--291},
	reportnumber = {IASSNS-HEP-98-15},
	title = {{Anti-de Sitter space and holography}},
	volume = {2},
	year = {1998}}

@article{Kibe:2021gtw,
    author = "Kibe, Tanay and Mandayam, Prabha and Mukhopadhyay, Ayan",
    title = "{Holographic spacetime, black holes and quantum error correcting codes: a review}",
    eprint = "2110.14669",
    archivePrefix = "arXiv",
    primaryClass = "hep-th",
    doi = "10.1140/epjc/s10052-022-10382-1",
    journal = "Eur. Phys. J. C",
    volume = "82",
    number = "5",
    pages = "463",
    year = "2022"
}

@article{Karch:2000ct,
    author = "Karch, Andreas and Randall, Lisa",
    editor = "Duff, Michael J. and Liu, J. T. and Lu, J.",
    title = "{Locally localized gravity}",
    eprint = "hep-th/0011156",
    archivePrefix = "arXiv",
    reportNumber = "MIT-CTP-3099",
    doi = "10.1088/1126-6708/2001/05/008",
    journal = "JHEP",
    volume = "05",
    pages = "008",
    year = "2001"
}

@article{DeWolfe:2001pq,
    author = "DeWolfe, Oliver and Freedman, Daniel Z. and Ooguri, Hirosi",
    title = "{Holography and defect conformal field theories}",
    eprint = "hep-th/0111135",
    archivePrefix = "arXiv",
    reportNumber = "CALT-68-2359, CITUSC-01-041, NSF-ITP-01-172, MIT-CTP-3212",
    doi = "10.1103/PhysRevD.66.025009",
    journal = "Phys. Rev. D",
    volume = "66",
    pages = "025009",
    year = "2002"
}

@article{Bachas:2020yxv,
    author = "Bachas, Constantin and Chapman, Shira and Ge, Dongsheng and Policastro, Giuseppe",
    title = "{Energy Reflection and Transmission at 2D Holographic Interfaces}",
    eprint = "2006.11333",
    archivePrefix = "arXiv",
    primaryClass = "hep-th",
    doi = "10.1103/PhysRevLett.125.231602",
    journal = "Phys. Rev. Lett.",
    volume = "125",
    number = "23",
    pages = "231602",
    year = "2020"
}

@article{Azeyanagi:2007qj,
    author = "Azeyanagi, Tatsuo and Karch, Andreas and Takayanagi, Tadashi and Thompson, Ethan G.",
    title = "{Holographic calculation of boundary entropy}",
    eprint = "0712.1850",
    archivePrefix = "arXiv",
    primaryClass = "hep-th",
    reportNumber = "KUNS-2114",
    doi = "10.1088/1126-6708/2008/03/054",
    journal = "JHEP",
    volume = "03",
    pages = "054",
    year = "2008"
}

@article{Afxonidis:2025jph,
    author = "Afxonidis, Evangelos and Carre{\~n}o Bolla, Ignacio and Hoyos, Carlos and Karch, Andreas",
    title = "{Connecting boundary entropy and effective central charge at holographic interfaces}",
    eprint = "2507.09171",
    archivePrefix = "arXiv",
    primaryClass = "hep-th",
    month = "7",
    year = "2025"
}

@article{Afxonidis:2024gne,
    author = "Afxonidis, Evangelos and Karch, Andreas and Murdia, Chitraang",
    title = "{The boundary entropy function for interface conformal field theories}",
    eprint = "2412.05381",
    archivePrefix = "arXiv",
    primaryClass = "hep-th",
    doi = "10.1007/JHEP07(2025)132",
    journal = "JHEP",
    volume = "07",
    pages = "132",
    year = "2025"
}

@article{Meineri:2019ycm,
    author = "Meineri, Marco and Penedones, Joao and Rousset, Antonin",
    title = "{Colliders and conformal interfaces}",
    eprint = "1904.10974",
    archivePrefix = "arXiv",
    primaryClass = "hep-th",
    doi = "10.1007/JHEP02(2020)138",
    journal = "JHEP",
    volume = "02",
    pages = "138",
    year = "2020"
}

@article{Henningson:1998gx,
	archiveprefix = {arXiv},
	author = {Henningson, M. and Skenderis, K.},
	doi = {10.1088/1126-6708/1998/07/023},
	eprint = {hep-th/9806087},
	journal = {JHEP},
	pages = {023},
	reportnumber = {CERN-TH-98-188, KUL-TF-98-21},
	title = {{The Holographic Weyl anomaly}},
	volume = {07},
	year = {1998}}

@article{Balasubramanian:1999re,
    author = "Balasubramanian, Vijay and Kraus, Per",
    title = "{A Stress tensor for Anti-de Sitter gravity}",
    eprint = "hep-th/9902121",
    archivePrefix = "arXiv",
    reportNumber = "HUTP-99-A002, EFI-99-6, NSF-ITP-98-132",
    doi = "10.1007/s002200050764",
    journal = "Commun. Math. Phys.",
    volume = "208",
    pages = "413--428",
    year = "1999"
}

@article{Wong:1994np,
    author = "Wong, E. and Affleck, I.",
    title = "{Tunneling in quantum wires: A Boundary conformal field theory approach}",
    eprint = "cond-mat/9311040",
    archivePrefix = "arXiv",
    doi = "10.1016/0550-3213(94)90479-0",
    journal = "Nucl. Phys. B",
    volume = "417",
    pages = "403--438",
    year = "1994"
}

@article{Oshikawa:1996dj,
    author = "Oshikawa, Masaki and Affleck, Ian",
    title = "{Boundary conformal field theory approach to the critical two-dimensional Ising model with a defect line}",
    eprint = "cond-mat/9612187",
    archivePrefix = "arXiv",
    doi = "10.1016/S0550-3213(97)00219-8",
    journal = "Nucl. Phys. B",
    volume = "495",
    pages = "533--582",
    year = "1997"
}

@article{Quella:2006de,
    author = "Quella, Thomas and Runkel, Ingo and Watts, Gerard M. T.",
    title = "{Reflection and transmission for conformal defects}",
    eprint = "hep-th/0611296",
    archivePrefix = "arXiv",
    reportNumber = "KCL-MTH-06-12, NSF-KITP-06-110",
    doi = "10.1088/1126-6708/2007/04/095",
    journal = "JHEP",
    volume = "04",
    pages = "095",
    year = "2007"
}

@article{Kimura:2015nka,
    author = "Kimura, Taro and Murata, Masaki",
    title = "{Transport Process in Multi-Junctions of Quantum Systems}",
    eprint = "1505.05275",
    archivePrefix = "arXiv",
    primaryClass = "hep-th",
    reportNumber = "HRI-ST-1504, RIKEN-MP-112, RIKEN-STAMP-7",
    doi = "10.1007/JHEP07(2015)072",
    journal = "JHEP",
    volume = "07",
    pages = "072",
    year = "2015"
}

@article{Billo:2016cpy,
    author = "Bill{\`o}, Marco and Gon{\c{c}}alves, Vasco and Lauria, Edoardo and Meineri, Marco",
    title = "{Defects in conformal field theory}",
    eprint = "1601.02883",
    archivePrefix = "arXiv",
    primaryClass = "hep-th",
    doi = "10.1007/JHEP04(2016)091",
    journal = "JHEP",
    volume = "04",
    pages = "091",
    year = "2016"
}

@article{Bousso:2015mna,
    author = "Bousso, Raphael and Fisher, Zachary and Leichenauer, Stefan and Wall, Aron C.",
    title = "{Quantum focusing conjecture}",
    eprint = "1506.02669",
    archivePrefix = "arXiv",
    primaryClass = "hep-th",
    doi = "10.1103/PhysRevD.93.064044",
    journal = "Phys. Rev. D",
    volume = "93",
    number = "6",
    pages = "064044",
    year = "2016"
}

@article{Wall:2017blw,
    author = "Wall, Aron C.",
    title = "{Lower Bound on the Energy Density in Classical and Quantum Field Theories}",
    eprint = "1701.03196",
    archivePrefix = "arXiv",
    primaryClass = "hep-th",
    doi = "10.1103/PhysRevLett.118.151601",
    journal = "Phys. Rev. Lett.",
    volume = "118",
    number = "15",
    pages = "151601",
    year = "2017"
}

@article{Kibe:2025cqc,
    author = "Kibe, Tanay and Roy, Pratik",
    title = "{Quantum null energy condition in quenched 2D CFTs}",
    eprint = "2503.17448",
    archivePrefix = "arXiv",
    primaryClass = "hep-th",
    doi = "10.1103/2p7v-f28f",
    journal = "Phys. Rev. D",
    volume = "111",
    number = "12",
    pages = "126009",
    year = "2025"
}

@article{Oshikawa:2005fh,
    author = "Oshikawa, Masaki and Chamon, Claudio and Affleck, Ian",
    title = "{Junctions of three quantum wires}",
    eprint = "cond-mat/0509675",
    archivePrefix = "arXiv",
    doi = "10.1088/1742-5468/2006/02/P02008",
    journal = "J. Stat. Mech.",
    volume = "0602",
    pages = "P02008",
    year = "2006"
}

@article{Chamon:2003tz,
    author = "Chamon, Claudio and Oshikawa, Masaki and Affleck, Ian",
    title = "{Junctions of three quantum wires and the dissipative Hofstadter model}",
    eprint = "cond-mat/0305121",
    archivePrefix = "arXiv",
    doi = "10.1103/PhysRevLett.91.206403",
    journal = "Phys. Rev. Lett.",
    volume = "91",
    pages = "206403",
    year = "2003"
}

@article{Gromov:2016umy,
    author = "Gromov, Andrey",
    title = "{Geometric Defects in Quantum Hall States}",
    eprint = "1604.03988",
    archivePrefix = "arXiv",
    primaryClass = "cond-mat.str-el",
    doi = "10.1103/PhysRevB.94.085116",
    journal = "Phys. Rev. B",
    volume = "94",
    number = "8",
    pages = "085116",
    year = "2016"
}

@article{fal1999topological,
  title={Topological defects and Goldstone excitations in domain walls between ferromagnetic quantum Hall effect liquids},
  author={Fal'ko, Vladimir I and Iordanskii, SV},
  journal={arXiv preprint cond-mat/9901053},
  year={1999}
}

@article{fendley1995exact,
  title={Exact conductance through point contacts in the $\nu$= 1/3 fractional quantum Hall effect},
  author={Fendley, P and Ludwig, AWW and Saleur, H},
  journal={Physical review letters},
  volume={74},
  number={15},
  pages={3005},
  year={1995},
  publisher={APS}
}

@article{Karch:2001cw,
    author = "Karch, Andreas and Randall, Lisa",
    title = "{Localized gravity in string theory}",
    eprint = "hep-th/0105108",
    archivePrefix = "arXiv",
    doi = "10.1103/PhysRevLett.87.061601",
    journal = "Phys. Rev. Lett.",
    volume = "87",
    pages = "061601",
    year = "2001"
}

@article{Bachas:2001vj,
    author = "Bachas, C. and de Boer, J. and Dijkgraaf, R. and Ooguri, H.",
    title = "{Permeable conformal walls and holography}",
    eprint = "hep-th/0111210",
    archivePrefix = "arXiv",
    reportNumber = "CALT-68-2361, CITUSC-01-045, ITFA-2001-33, LPTENS-01-42",
    doi = "10.1088/1126-6708/2002/06/027",
    journal = "JHEP",
    volume = "06",
    pages = "027",
    year = "2002"
}

@article{Israel:1966rt,
    author = "Israel, W.",
    title = "{Singular hypersurfaces and thin shells in general relativity}",
    doi = "10.1007/BF02710419",
    journal = "Nuovo Cim. B",
    volume = "44S10",
    pages = "1",
    year = "1966",
    note = "[Erratum: Nuovo Cim.B 48, 463 (1967)]"
}

@article{Banados:1998gg,
    author = "Banados, Maximo",
    editor = "Falomir, H. and Gamboa Saravi, R. E. and Schaposnik, F. A.",
    title = "{Three-dimensional quantum geometry and black holes}",
    eprint = "hep-th/9901148",
    archivePrefix = "arXiv",
    doi = "10.1063/1.59661",
    journal = "AIP Conf. Proc.",
    volume = "484",
    number = "1",
    pages = "147--169",
    year = "1999"
}

@article{Son:2002sd,
    author = "Son, Dam T. and Starinets, Andrei O.",
    title = "{Minkowski space correlators in AdS / CFT correspondence: Recipe and applications}",
    eprint = "hep-th/0205051",
    archivePrefix = "arXiv",
    reportNumber = "INT-PUB-02-34",
    doi = "10.1088/1126-6708/2002/09/042",
    journal = "JHEP",
    volume = "09",
    pages = "042",
    year = "2002"
}

@article{Herzog:2002pc,
    author = "Herzog, C. P. and Son, D. T.",
    title = "{Schwinger-Keldysh propagators from AdS/CFT correspondence}",
    eprint = "hep-th/0212072",
    archivePrefix = "arXiv",
    reportNumber = "NSF-ITP-02-175, INT-PUB-02-53",
    doi = "10.1088/1126-6708/2003/03/046",
    journal = "JHEP",
    volume = "03",
    pages = "046",
    year = "2003"
}

@article{Skenderis:2008dg,
    author = "Skenderis, Kostas and van Rees, Balt C.",
    title = "{Real-time gauge/gravity duality: Prescription, Renormalization and Examples}",
    eprint = "0812.2909",
    archivePrefix = "arXiv",
    primaryClass = "hep-th",
    reportNumber = "ITFA-2008-50",
    doi = "10.1088/1126-6708/2009/05/085",
    journal = "JHEP",
    volume = "05",
    pages = "085",
    year = "2009"
}

@article{Bianchi:2015liz,
    author = "Bianchi, Lorenzo and Meineri, Marco and Myers, Robert C. and Smolkin, Michael",
    title = "{R{\'e}nyi entropy and conformal defects}",
    eprint = "1511.06713",
    archivePrefix = "arXiv",
    primaryClass = "hep-th",
    reportNumber = "DESY-15-229",
    doi = "10.1007/JHEP07(2016)076",
    journal = "JHEP",
    volume = "07",
    pages = "076",
    year = "2016"
}

@article{harlow2018tasi,
	author = "Harlow, Daniel",
    title = "{TASI Lectures on the Emergence of Bulk Physics in AdS/CFT}",
    eprint = "1802.01040",
    archivePrefix = "arXiv",
    primaryClass = "hep-th",
    doi = "10.22323/1.305.0002",
    journal = "PoS",
    volume = "TASI2017",
    pages = "002",
    year = "2018"}

@article{Jahn:2021uqr,
	archiveprefix = {arXiv},
	author = {Jahn, Alexander and Eisert, Jens},
	doi = {10.1088/2058-9565/ac0293},
	eprint = {2102.02619},
	journal = {Quantum Sci. Technol.},
	number = {3},
	pages = {033002},
	primaryclass = {quant-ph},
	title = {{Holographic tensor network models and quantum error correction: a topical review}},
	volume = {6},
	year = {2021}}

@article{Chen:2021lnq,
	archiveprefix = {arXiv},
	author = {Chen, Bowen and Czech, Bartlomiej and Wang, Zi-zhi},
	doi = {10.1088/1361-6633/ac51b5},
	eprint = {2108.09188},
	journal = {Rept. Prog. Phys.},
	number = {4},
	pages = {046001},
	primaryclass = {hep-th},
	title = {{Quantum information in holographic duality}},
	volume = {85},
	year = {2022}}

@article{deHaro:2000vlm,
    author = "de Haro, Sebastian and Solodukhin, Sergey N. and Skenderis, Kostas",
    title = "{Holographic reconstruction of space-time and renormalization in the AdS / CFT correspondence}",
    eprint = "hep-th/0002230",
    archivePrefix = "arXiv",
    reportNumber = "SPIN-2000-05, ITP-UU-00-03, PUTP-1921",
    doi = "10.1007/s002200100381",
    journal = "Commun. Math. Phys.",
    volume = "217",
    pages = "595--622",
    year = "2001"
}

@article{Iizuka:2025xnd,
    author = "Iizuka, Norihiro and Ishibashi, Akihiro and Maeda, Kengo and Nakayama, Haruki and Nishioka, Tatsuma",
    title = "{Energy Conditions and Quantum Information}",
    eprint = "2509.01286",
    archivePrefix = "arXiv",
    primaryClass = "hep-th",
    reportNumber = "NU-QG-10, OU-HET-1283",
    month = "9",
    year = "2025"
}

@misc{supp,
  howpublished={See Supplemental Material at \url{https://doi.org/10.1103/d76f-x3c7}}
}

\end{document}